%% file: ICRC2025-Mizar.tex
\documentclass[a4paper,11pt]{article}
\usepackage{pos}
\usepackage{xurl}
\usepackage{hyperref}
\usepackage{natbib}
\usepackage{graphicx}
\usepackage{float}

\title{Performance results of the first version of the MIZAR ASIC for the PBR mission}

\author*[a,b]{M. Bertaina}
\author[a,b]{P.~A. Palmieri}
\author[b]{M. Bargelli}
\author[a,b]{M.~D. Da Rocha Rolo}
\author[a]{G. Dellacasa}
\author[a]{A. Di Salvo}
\author[a]{S. Garbolino}
\author[a]{G. Mazza}
\author[a]{M. Mignone}
\author[a,b]{A. Rivetti}
\author[c]{G. Traversi}
\author[b]{E. Trossarello}
\author[a]{R. Wheadon}
\author[a,b]{S.~C. Zugravel}

\affiliation[a]{INFN Section of Turin,
Via P. Giuria 1, 10125 Turin, Italy}

\affiliation[b]{University of Turin, Department of Physics,
  V. P. Giuria 1, 10125 Turin, Italy}

\affiliation[c]{University of Bergamo, Viale Marconi 5, 24044 Dalmine, Italy}

\onbehalf{for the JEM-EUSO Collaboration\\[-1mm]{\normalsize \normalfont (a complete list of authors can be found at the end of the proceedings)}}

\emailAdd{bertaina@to.infn.it}

\abstract{The Multi-channel Intergrated Zone-sampling Analogue-memory based Readout (MIZAR) ASIC is a new type of front-end electronics which has been developed for the detection of the optical Cherenkov signals by Extensive Air Showers directly observed from sub-orbital and orbital altitudes. It sets the stage for a new generation of low-power consuming 64-channel Application-Specific Integrated Circuit (ASIC). The ASIC is implemented in a commercial 65 nm CMOS technology to readout an 8$\times$8 matrix of Silicon Photo-Multipliers (SiPMs). The event is recorded at channel level in an array of 256 cells, each one equipped with an analogue memory, a 12-bits Wilkinson Analog-to-Digital Converter (ADC) and latches running at a sampling rate of 200 MS/s. The converted data are sent off-chip to a Field Programmable Gate Array (FPGA) which controls the digital end-of-column logic of the ASIC and implements the trigger logic. The goal is to employ it for the first time on the POEMMA Balloon with Radio (PBR) NASA mission which is devoted to measure Ultra-High Energy Cosmic Rays (UHECRs) and perform neutrino astronomy from stratospheric altitudes through the detection of atmospheric Cherenkov light. The first version of the MIZAR ASIC has been sent to production and it is now under test at INFN Torino. The results of the preliminary tests related to the characterization of the chip are presented.}

\ConferenceLogo{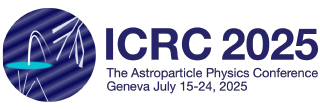}

\FullConference{39th International Cosmic Ray Conference (ICRC2025)\\
 15–24 July 2025\\
Geneva, Switzerland\\}

\begin{document}
\maketitle

\section{Introduction}
\label{sec:intro}

Detecting Cherenkov light emitted by Extensive Air Showers (EAS) initiated by Ultra-High Energy Cosmic Rays (UHECRs) or neutrinos presents specific challenges, particularly when observations are performed from near-space platforms, such as low Earth orbit ($\sim$500 km) or sub-orbital altitudes (30–40 km). Simulations suggest that the Cherenkov signals generated under these conditions can persist for several tens of nanoseconds, depending on the angular separation between the shower axis and the telescope’s line of sight. This implies that the Front-End Electronics (FEE) must be capable of time resolutions of around 10 ns or better to adequately capture the signal’s temporal profile. In addition to high time resolution, waveform acquisition capabilities become essential to distinguish EAS-related signals from other spurious events, such as direct cosmic ray hits on the detector. 

Responding to these experimental demands, INFN Torino initiated the design of a new generation of Application-Specific Integrated Circuits (ASICs). This ASIC is tailored to meet the stringent timing and triggering requirements for high-energy astroparticle detection. The development is closely aligned with recent, ongoing, and future missions focused on UHECR and neutrino detection, including stratospheric balloon experiments like EUSO-SPB2~\cite{Eser_2023} and space-based concepts such as Terzina~\cite{terzina}, POEMMA~\cite{poemma} and M-EUSO~\cite{Zbigniew_ICRC2025}. A preliminary overview of the ASIC architecture can be found in~\cite{twepp}, with more recent iterations detailed in~\cite{tredi}. In this context, the chip’s performance is being evaluated as part of the POEMMA-Balloon-Radio (PBR) mission~~\cite{Eser:2025/T}.

The PBR mission~\cite{Eser:2025/T} is the next balloon-borne mission of the JEM-EUSO program after the EUSO-SPB2 experiment~\cite{Eser_2023}, which flew aboard a NASA Super Pressure Balloon launched on May 13, 2023, from Wanaka, New Zealand. EUSO-SPB2 employed a dual-instrument payload: a fluorescence telescope with a MAPMT focal plane aimed at nadir to observe fluorescence from EeV-scale EAS, and a Cherenkov telescope (CTel) equipped with a SiPM camera to detect Cherenkov light from PeV-scale EAS, including those from tau neutrino decays.

The PBR concept builds upon this heritage, however, with some significant differences. The major component of PBR
are one large (1-meter diameter), tiltable telescope housing a hybrid focal surface for both fluorescence and Cherenkov measurements. The payload includes also a Radio Instrument (RI) of two antennas mounted beneath the telescope. The optical component is designed to produce two distinct focal spots on the camera for the Cherenkov measurements. More details can be found in~\cite{eser_icrc2025}. This optical design enhances event discrimination for Cherenkov measurements by allowing the system to distinguish between single-spot direct hits from cosmic rays and dual-spot patterns from off-axis background light, effectively reducing false triggers and improving the signal-to-noise ratio.

At the heart of the Cherenkov component of the instrument is a high-speed focal plane camera, consisting of 512 Silicon Photomultiplier (SiPM) pixels (Hamamatsu S14521-6050AN-04). The camera is optimized for detecting nanosecond-scale optical pulses with a 10 ns integration time and a 512-frame analog memory for signal buffering. The readout system is centered around a dedicated front-end ASIC that performs fast waveform sampling, analog-to-digital conversion, and data derandomization. This architecture enables the precise reconstruction of temporal signal profiles and supports high-throughput event capture with minimal dead time.

The telescope’s field of view spans 6.4$^{\circ}$ in zenith and 12.8$^{\circ}$ in azimuth. During flight, it can be pointed across a range of elevation angles—from the horizontal plane to 10$^{\circ}$ below the Earth’s limb—allowing for diverse observational strategies. These include detecting above-the-limb Cherenkov light from UHECR-induced EAS and potentially observing tau-lepton decays resulting from Earth-skimming UHE neutrino interactions.

\section{The Front end Electronics}
\label{sec:fee}

\begin{figure}[htbp]
    	\centering
    	\includegraphics[width=0.9\linewidth]{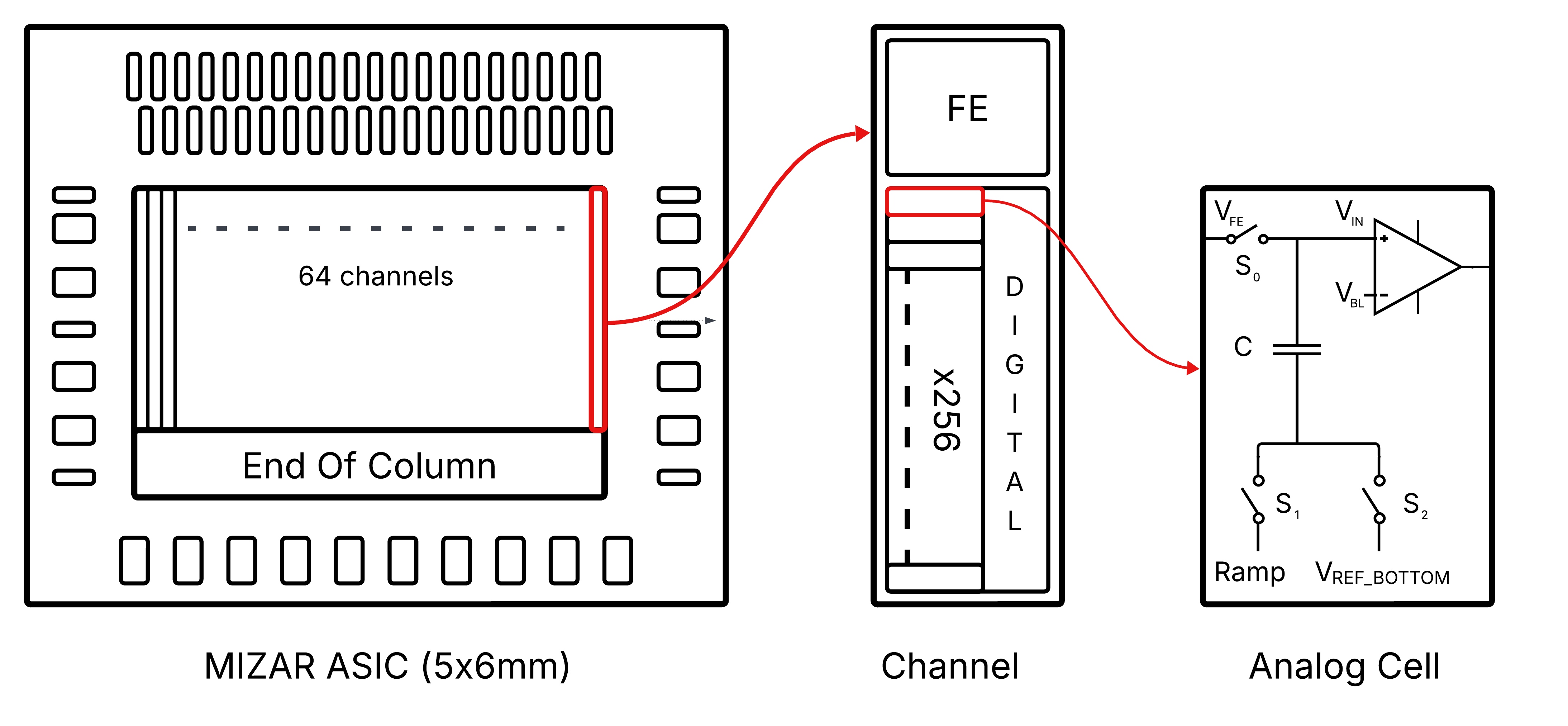}	
    	\caption{Block diagram of the MIZAR architecture.}
    	\label{fig:complete_mizar_diagram}
    \end{figure}

The custom-designed ASIC, realized in a commercial 65-nm CMOS process and operating at a 1.2 V supply, offers a highly energy-efficient solution for applications requiring low-power, high-performance signal processing. Specifically developed to handle 8$\times$8 arrays of SiPMs, it provides 64 parallel channels capable of waveform sampling for accurate reconstruction of incoming signals.

Figure \ref{fig:complete_mizar_diagram} illustrates a block diagram of the MIZAR architecture. Each channel features a Front-End (FE) analog block which generates a signal distributed among an array of 256 cells, equipped with analog memories and sampling the input at the nominal frequency of 200 MHz. The memory architecture integrates a storage capacitor and a single-slope ADC per cell, a configuration that simplifies hardware complexity while allowing parallel digitization of all stored samples. This approach significantly reduces system dead time without compromising precision. To further optimize power usage, digitization is decoupled from sampling and is triggered only upon event validation through internal or external trigger signals. The ADC resolution is configurable from 8 to 12 bits, enabling lower-resolution operation for background monitoring, which decreases digitization overhead when fine detail is unnecessary. The analog memory can function as a monolithic unit or be segmented into 32- or 64-cell buffers, enabling multi-buffer acquisition. This buffering allows data from one event to be digitized while the system remains available for subsequent triggers, thus minimizing event losses due to readout latency. 

    \begin{figure}[htbp]
    	\centering
    	\includegraphics[width=\linewidth]{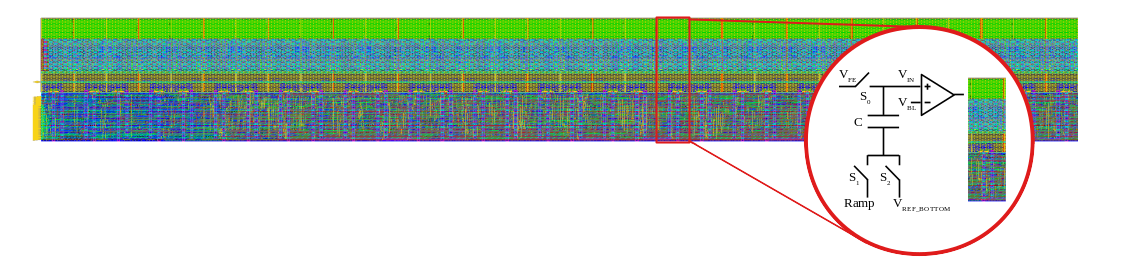}	
    	\caption{Layout of a section made of 32 cells. In the round area a single cell is magnified next to the analog schematic.}
    	\label{fig:section_layout}
    \end{figure}

The chip integrates both analog and digital circuitry on a single die using a digital-on-top layout strategy. Figure \ref{fig:section_layout} shows the physical layout of a section as example of this approach. With a compact footprint of 6$\times$5 mm² and a target power consumption of 5 mW per channel, the ASIC meets the stringent requirements of future space-based and sub-orbital UHECR and neutrino observatories.

\begin{figure}[htbp]
    	\centering
    	\includegraphics[width=0.6\linewidth]{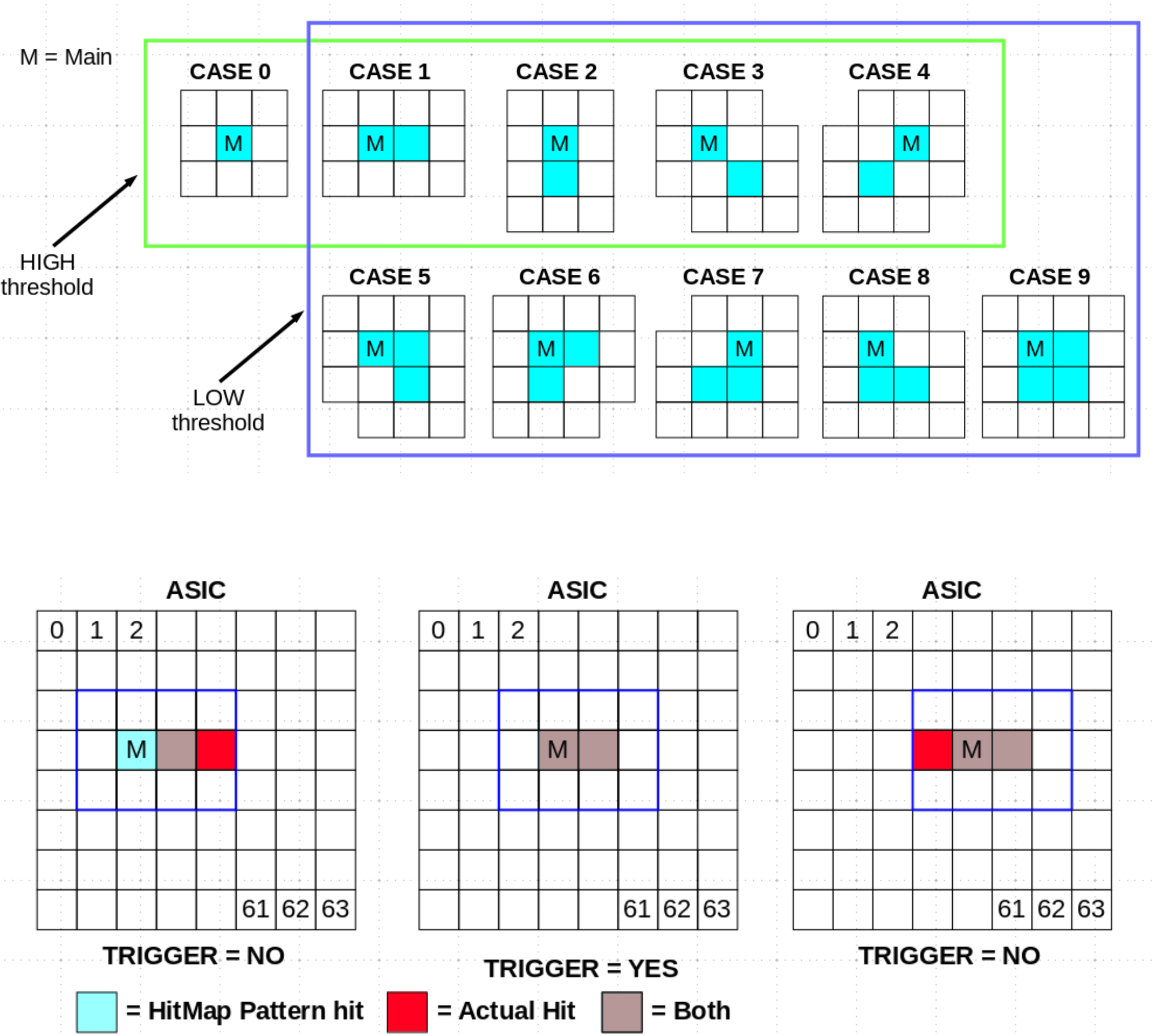}	
    	\caption{[Top] Hitmap validator pattern case: various patterns are searched by the FPGA to determine whether data is accepted or rejected. [Bottom] Pattern matching algorithm: each pixel is classified as main (M), and the pattern fingerprint is verified. This procedure is carried out in parallel for every pixel and each pattern case~\cite{Bertaina:2023}.}
    	\label{fig:pattern_recognition}
    \end{figure}

The trigger and readout system is designed to efficiently identify relevant events while minimizing unnecessary data throughput (see Figure~\ref{fig:pattern_recognition}). Each channel includes two programmable thresholds: a high threshold, used to flag events where most of the signal is localized in a single pixel, and a low threshold, which captures events distributed over neighboring pixels. When the signal exceeds the low threshold, a programmable counter initiates a validation window. If the signal also crosses the high threshold within this interval, only the high-threshold trigger is forwarded to the FPGA; otherwise, the low-threshold trigger is sent. The FPGA then evaluates the 64-bit hitmap from each ASIC, which encodes the active channels, and decides whether to accept the event for readout. The FPGA follows three decision paths:
    \begin{enumerate}
        \item \textit{Event Acceptance:} If the high threshold is exceeded in one or two adjacent pixels, or if the low threshold is exceeded in two to four neighboring pixels, the event is flagged for readout.
    	\item \textit{Event Rejection:} If the hitmap lacks sufficient spatial correlation, the event is discarded. In cases where a single edge pixel is triggered, the FPGA may check adjacent ASICs for extended event signatures before rejecting the event.
    	\item \textit{No Response:} If the FPGA fails to respond within a defined time window ($\Delta t_c \sim80$ ns), the system resets and resumes monitoring new triggers.
    \end{enumerate}

Upon event acceptance, digitization is centered on the time at which the signal crosses threshold ($t_S$). The memory architecture supports flexible segmentation from 32 to 256 cells per channel, with resolutions selectable between 8 and 12 bits. A typical setup uses 8 independent 32-cell blocks, capturing a 160 ns time window with 12-bit resolution. The sampled waveform consists of 32 points spaced by $T_{clk}$ = 5 ns, covering the interval from $t_S - 80$ ns to $t_S + 80$ ns.

The digitization of an individual event requires a maximum of 20.5 $\mu$s ($2^{12} \times T_{clk}$). While one block undergoes conversion, the remaining seven are still available for incoming events. If all buffers are occupied, only the hitmap is sent, and new data cannot be recorded until space is freed. The data output per channel comprises a 48-bit header and 386 bits of digitized waveform, totaling 434 bits. For the entire ASIC, this corresponds to 27,776 bits per event. Transmission is handled by a DDR serializer operating at 400 MHz, with a full ASIC readout taking 34.7 $\mu$s. Thanks to eight parallel conversion chains and assuming a single FPGA can manage five ASICs, the system remains responsive even under high trigger rates. Deadtime occurs only when all buffers are full simultaneously, a scenario mitigated by the memory segmentation and derandomization scheme.

\section{Results of the preliminary tests on the MIZAR ASIC}
\label{sec:tests}
In the first quarter of the year a batch of 200 MIZAR ASICs have been delivered by the foundry and six samples have been bonded on the Front-End Boards (FEBs) received in the same period. In the following a description of the test setup and the results are provided. The test campaign considered the six available chips whose functionalities have been under investigation.

\subsection{Setup description}

\begin{figure}[H]
    \centering
    
    \includegraphics[width=\linewidth]{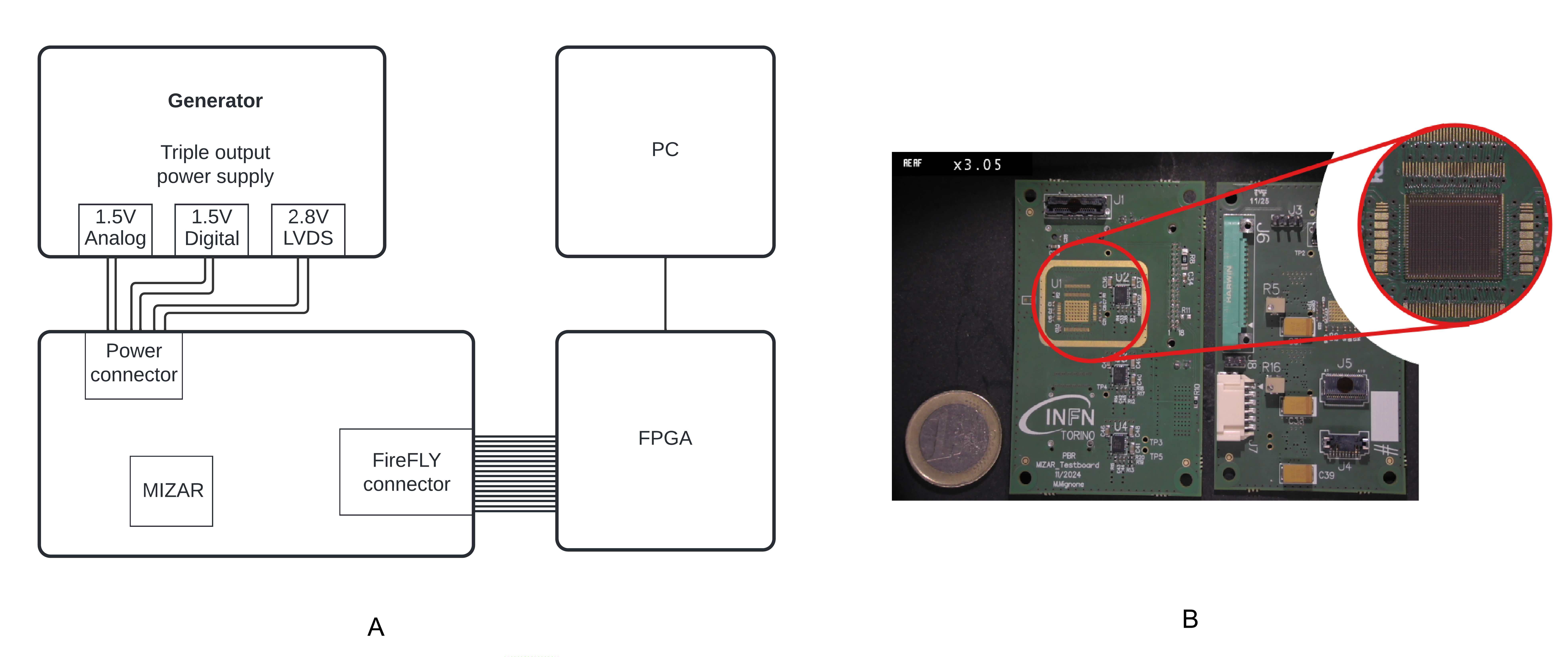}	
    \caption{A. Block diagram of testing setup. B. FEB and MIZAR bonded on the board}
    \label{fig:diagram_board}
\end{figure}

Figure \ref{fig:diagram_board}, A depicts a block diagram representation of the test setup used in laboratory. Each MIZAR ASIC has been bonded on a FEB which is connected to a power supply. The generator was set to 1.5 V for both analog and digital blocks and to 2.8 V for the Low-Voltage Differential Signaling (LVDS) lines. The board itself then provides the three power domains to the chips through its voltage regulators. Figure \ref{fig:diagram_board}, B shows front and back sides of the FEB while the magnified area is a close up of the bonded ASIC. The MIZAR chip is interfaced to the FPGA with a FireFly cable to ensure the configuration and the communication.
All chips are active and the power consumption values are consistent with the < 5 mW/channel requirement expected from the simulations. 

\subsection{SPI and ASIC configuration}

\begin{table}[htbp]
\begin{center}
\begin{tabular}{|ccc|}
\hline
Parameter & Value & Description\\
\hline \hline

\hline
High VTH & 15 & Channel high threshold \\
\hline
Low VTH & 15 & Channel low threshold \\
\hline
CTR\_TP\_VB & 250 $\mu A$ & Test pulse amplitude\\
\hline
CTR\_MIR\_VB & 320 $nA$ & Current supplied by the ramp generator per unit time\\
\hline
CTR\_LO\_VTH\_GBL & 900 $mV$ & High threshold baseline\\
\hline
CTR\_HI\_VTH\_GBL & 900 $mV$ & Low threshold baseline \\
\hline
VTH\_LSB & 1 $mV$ & LSB value \\
\hline
GCR\_DC\_COUPLING & on & DC coupling\\
\hline
TP sequence & 20 $ns$ & Test pulse duration \\
\hline

\end{tabular}
\vspace{10pt}
\caption{Parameters used for data acquisition.}\label{tab:registers}
\end{center}
\end{table}

\noindent
Table \ref{tab:registers} reports the values of the main parameters used during data acquisition. The High and Low thresholds span a range determined by the VTH\_GBL and VTH\_LSB registers, according to the relation VTH\_GBL - n $\times$ VTH\_LSB, with n ranging from 0 to 31. A value of n = 31 corresponds to the baseline set by the CTR\_LO\_VTH\_GBL and CTR\_HI\_VTH\_GBL registers, while n = 0 represents the baseline reduced by 31 LSB steps. The test pulse injected into each channel is characterized by the CTR\_TP\_VB register, which sets its amplitude, and by the TP sequence section, which defines its duration. The configuration also includes the CTR\_MIR\_VB register, which controls the slope of the ramp used in the ADC conversion by regulating the current supplied over time. The current can be programmed to vary from 40 $nA$ to 640 $nA$ in discrete steps of 40 $nA$.Throughout the tests, DC coupling was used by enabling the GCR\_DC\_COUPLING register.

\subsection{Test pulse}
The FE of each channel can be configured to process the input signal coming from the SiPM or to generate a Test Pulse (TP). Latter was enabled to study both the hitmap generation and the data serialization. The TP provided to each channel was enabled via Serial Peripheral Interface (SPI) by writing the PCR\_CAL\_MODE register and its amplitude and duration were programmable as well. In this test mode the FPGA was programmed to force the ASICs to provide the raw data of the hitmap and the converted data. The whole group of available bonded chips successfully sent out the required information, proving the testability of each MIZAR ASIC.

 \begin{figure}[htbp]
    	\centering
    	\includegraphics[width=0.6\linewidth]{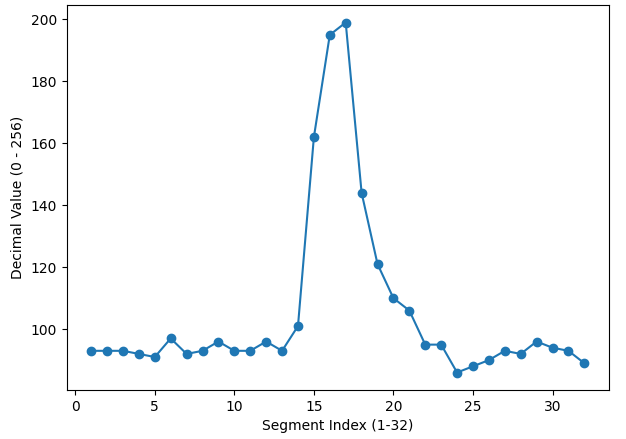}	
    	\caption{Test pulse acquisition on channel 5, chip 1. 8 bit configuration, 32 cells partitioning.}
    	\label{fig:test_pulse}
    \end{figure}

Figure \ref{fig:test_pulse} reports a typical pulse internally generated by the FE, converted by the Wilkinson ADCs and serialized by the digital logic of the end of column. The test was realized at the sampling frequency of 100 MHz using the 32-cells partition and the 8 bits resolution. These results were carried out without any baseline correction yet. In addition, a preliminary characterization of the preamplifier gain response was conducted by varying the 4 bit GCR\_GAIN\_CTR register settings. The measured gain behavior demonstrated consistency with the design specifications. Finally, a preliminary study on the signals has identified a possible fixed pattern noise, but this feature is currently under investigation since it requires more data acquisitions.

\section{Conclusions}
\label{sec:conclusions}
A new type of FEE for the detection of the optical Cherenkov signals by EAS directly observed from sub-orbital and orbital altitudes has been described. It is based on a new type of ASIC which is specifically developed for space applications with low power consumption ($\sim$5 mW/channel) and sufficiently radiation hard. The ASIC’s 200 MHz clock frequency enables a precise waveform
sampling to clearly recognize an EAS Cherenkov signal from other types of sources such as direct cosmic ray hits, noise spikes or atmospheric events. The developed trigger logic is capable of
reducing the fake event rates and the data overhead in order to downlink all the events of interest. The hitmaps also allow the possibility of triggering on signals generated by a bi-focal system. This is considered to be another key way to uniquely recognize an EAS event. The first version of the MIZAR ASIC has been produced and it is currently under tests. Preliminary results indicate a performance in line with the expected basic characteristics of the chip. A more detailed evaluation will follow in the forthcoming months. The MIZAR ASIC is expected to fly on board the PBR mission.

\section{Acknowledgments}
\label{sec:acknowledgments}
The FEE described in this paper is part of the development performed within ASI-INFN agreement for EUSO-SPB2 n.2021-8-HH.0 and its amendments. We thank the JEM-EUSO, NUSES and POEMMA collaborations for the very fruitful discussions which inspired the design.

\bibliography{references}


\input{JEM-EUSO_Authors_July2025}

\end{document}

%% file: JEM-EUSO_Authors_July2025.tex
    \newpage
{\Large\bf Full Authors list: The JEM-EUSO Collaboration}

\begin{sloppypar}
{\small \noindent
M.~Abdullahi$^{ep,er}$              
M.~Abrate$^{ek,el}$,                
J.H.~Adams Jr.$^{ld}$,              
D.~Allard$^{cb}$,                   
P.~Alldredge$^{ld}$,                
R.~Aloisio$^{ep,er}$,               
R.~Ammendola$^{ei}$,                
A.~Anastasio$^{ef}$,                
L.~Anchordoqui$^{le}$,              
V.~Andreoli$^{ek,el}$
% ...
}
\end{sloppypar}
\vspace*{.3cm}

{ \footnotesize
\noindent
%
% ...
$^{cb}$ Universit\'e de Paris, CNRS, AstroParticule et Cosmologie, F-75013 Paris, France\\
% ...
$^{ef}$ Istituto Nazionale di Fisica Nucleare - Sezione di Napoli, Italy\\
% ...
$^{ei}$ Istituto Nazionale di Fisica Nucleare - Sezione di Roma Tor Vergata, Italy\\
% ...
$^{ek}$ Istituto Nazionale di Fisica Nucleare - Sezione di Torino, Italy\\
$^{el}$ Dipartimento di Fisica, Universit\`a di Torino, Italy\\
% ...
$^{ep}$ Gran Sasso Science Institute, L'Aquila, Italy\\
$^{er}$ INFN, Gran Sasso National Laboratory, Assergi (AQ), I-67100, Italy\\
% ...
$^{ld}$ University of Alabama in Huntsville, Huntsville, AL, USA\\
$^{le}$ Lehman College, City University of New York (CUNY), NY, USA\\
% ...
}

%% file: ICRC2025-Mizar.bbl
\providecommand{\href}[2]{#2}\begingroup\raggedright\begin{thebibliography}{1}

\bibitem{Eser_2023}
J.~Eser et~al., \emph{{Overview and First Results of EUSO-SPB2}},  in
  \emph{Proceedings of 38th International Cosmic Ray Conference —
  PoS(ICRC2023)}, ICRC2023, p.~397, Sissa Medialab, Aug., 2023,
  \href{https://doi.org/10.22323/1.444.0397}{DOI}.

\bibitem{terzina}
L.~Burmistrov et~al., \emph{{Terzina on board NUSES: A pathfinder for EAS
  Cherenkov Light Detection from space}},
  \href{https://doi.org/10.48550/arXiv.2304.11992}{\emph{EPJ Web of
  Conferences} {\bfseries 283} (2023) 06006}
  [\href{https://arxiv.org/abs/2304.11992}{{\ttfamily 2304.11992}}].

\bibitem{poemma}
A.~Olinto et~al., \emph{{The POEMMA (Probe of Extreme Multi-Messenger
  Astrophysics) observatory}},
  \href{https://doi.org/10.1088/1475-7516/2021/06/007}{\emph{JCAP} {\bfseries
  2021} (2021) 007} [\href{https://arxiv.org/abs/2012.07945}{{\ttfamily
  2012.07945}}].

\bibitem{Zbigniew_ICRC2025}
Z.~Plebaniak et~al., \emph{{From Ground to Space: An Overview of the JEM-EUSO
  Program for the Study of UHECRs and Astrophysical Neutrinos}}, {\emph{This
  Conference Porceedings} {\bfseries ICRC2025} (2025) }.

\bibitem{twepp}
S.~Tedesco et~al., \emph{{A 64-channel waveform sampling ASIC for SiPM in
  space-born applications}},
  \href{https://doi.org/10.1088/1748-0221/18/02/C02022}{\emph{Journal of
  Instrumentation} {\bfseries 18} (2023) C02022}.

\bibitem{tredi}
A.~Di~Salvo, \emph{{A low-power, 64-channel ASIC for space applications for
  Cherenkov radiation detection}},  in \emph{TREDI Workshop (Trento)}, 2023,
  \href{https://indico.cern.ch/event/1223972/timetable/?view=nicecompact}{https://indico.cern.ch/event/1223972/timetable/?view=nicecompact}.

\bibitem{Eser:2025/T}
J.~Eser et~al., \emph{{POEMMA-Balloon with Radio, towards a space-based
  Multi-Messenger Observatory}},
  \href{https://doi.org/10.22323/1.484.0061}{\emph{PoS} {\bfseries UHECR2024}
  (2025) 061}.

\bibitem{eser_icrc2025}
J.~Eser et~al., \emph{{POEMMA-Balloon with Radio: An Overview}}, {\emph{This
  Conf. Proceedings} {\bfseries ICRC2025} (2025) }.

\bibitem{Bertaina:2023}
M.~Bertaina et~al., \emph{{A new front end electronics for the detection of the
  optical Cherenkov signals by Extensive Air Showers directly observed from
  sub-orbital and orbital altitudes}},
  \href{https://doi.org/10.22323/1.444.0311}{\emph{PoS} {\bfseries ICRC2023}
  (2023) 311}.

\end{thebibliography}\endgroup
